\documentclass[aps,showpacs,showkeys]{revtex4}
\usepackage{amsmath,amsthm,amssymb,epsfig,alltt}

\begin{document}

\def\a{\alpha}
\def\b{\beta}
\def\d{{\delta}}
\def\l{\lambda}
\def\e{\epsilon}
\def\p{\partial}
\def\m{\mu}
\def\n{\nu}
\def\t{\tau}
\def\th{\theta}
\def\s{\sigma}
\def\g{\gamma}
\def\o{\omega}
\def\r{\rho}
\def\z{\zeta}
\def\D{\Delta}
\def\half{\frac{1}{2}}
\def\hatt{{\hat t}}
\def\hatx{{\hat x}}
\def\hatp{{\hat p}}
\def\hatX{{\hat X}}
\def\hatY{{\hat Y}}
\def\hatP{{\hat P}}
\def\haty{{\hat y}}
\def\whatX{{\widehat{X}}}
\def\whata{{\widehat{\alpha}}}
\def\whatb{{\widehat{\beta}}}
\def\whatV{{\widehat{V}}}
\def\hatth{{\hat \theta}}
\def\hatta{{\hat \tau}}
\def\hatrh{{\hat \rho}}
\def\hatva{{\hat \varphi}}
\def\barx{{\bar x}}
\def\bary{{\bar y}}
\def\barz{{\bar z}}
\def\baro{{\bar \omega}}
\def\barpsi{{\bar \psi}}
\def\sp{\sigma^\prime}
\def\nn{\nonumber}
\def\cb{{\cal B}}
\def\2pap{2\pi\alpha^\prime}
\def\wideA{\widehat{A}}
\def\wideF{\widehat{F}}
\def\beq{\begin{eqnarray}}
 \def\eeq{\end{eqnarray}}
 \def\4pap{4\pi\a^\prime}
 \def\op{\omega^\prime}
 \def\xp{{x^\prime}}
 \def\sp{{\s^\prime}}
 \def\ap{{\a^\prime}}
 \def\tp{{\t^\prime}}
 \def\zp{{z^\prime}}
 \def\xpp{x^{\prime\prime}}
 \def\xppp{x^{\prime\prime\prime}}
 \def\barxp{{\bar x}^\prime}
 \def\barxpp{{\bar x}^{\prime\prime}}
 \def\barxppp{{\bar x}^{\prime\prime\prime}}
 \def\zetap{{\zeta^\prime}}
 \def\barchi{{\bar \chi}}
 \def\baro{{\bar \omega}}
 \def\bpsi{{\bar \psi}}
 \def\barg{{\bar g}}
 \def\barz{{\bar z}}
 \def\bareta{{\bar \eta}}
 \def\ta{{\tilde \a}}
 \def\tb{{\tilde \b}}
 \def\tc{{\tilde c}}
 \def\tz{{\tilde z}}
 \def\tJ{{\tilde J}}
 \def\tpsi{\tilde{\psi}}
 \def\tal{{\tilde \alpha}}
 \def\tbe{{\tilde \beta}}
 \def\tga{{\tilde \gamma}}
 \def\tchi{{\tilde{\chi}}}
 \def\barth{{\bar \theta}}
 \def\bareta{{\bar \eta}}
 \def\barom{{\bar \omega}}
 \def\bole{{\boldsymbol \epsilon}}
 \def\bolth{{\boldsymbol \theta}}
 \def\bomega{{\boldsymbol \omega}}
 \def\bolmu{{\boldsymbol \mu}}
 \def\bolal{{\boldsymbol \alpha}}
 \def\bolbe{{\boldsymbol \beta}}
 \def\bolL{{\boldsymbol  L}}
 \def\bolX{{\boldsymbol X}}
 \def\boln{{\boldsymbol n}}
 \def\bolp{{\boldsymbol p}}
 \def\bolx{{\boldsymbol x}}
 \def\bols{{\boldsymbol s}}
 \def\bolS{{\boldsymbol S}}
 \def\bola{{\boldsymbol a}}
 \def\bolA{{\boldsymbol A}}
 \def\bolJ{{\boldsymbol J}}
 \def\tr{{\rm tr}}

\setcounter{page}{1}
\title[]{Entanglement Entropy for Open Bosonic Strings on $Dp$-branes
}

\author{Taejin Lee}
\email{taejin@kangwon.ac.kr}
\affiliation{Department of Physics, Kangwon National University, Chuncheon 24341 Korea}

\date{\today }

\begin{abstract}
We study the entanglement entropy for open bosonic strings on multiple $Dp$-branes by using the covariant 
open string field theory. Choosing one of the spatial coordinates which are tangential to the 
hyperplane on which $Dp$-branes are located, we divide the hyperplane into two halves. By using the string wavefunction 
in the Fock space representation, we evaluate the entanglement entropy. The entanglement 
entropy is found to be proportional to the area of $(p-1)$-dimensional boundary of the bipartite hyperplanes and 
divergent in the ultraviolet (UV) region  as well as in the infrared (IR) region. 
However, the leading divergences are mainly due to tachyon contributions to 
the entanglement entropy, which may be absent in  
supersymmetric string theories. Apart from the divergences thanks to tachyons, the 
entanglement entropy for open bosonic strings on $Dp$-branes is finite for $2 \le p \le d_{\text{critical}} -2$ and logarithmically divergent for $p =1, d_{\text{critical}}-1$.

\end{abstract}




\maketitle

\section{Introduction}

The entanglement entropy \cite{Eisert2010}, which is also termed as geometric entropy \cite{Bombelli1986,Srednicki1993,Holzhey94LW,Callan1994W,Dowker1994,Larsen1994W,Casini2004}, 
has been a focal point for recent developments which cover a wide range of theoretical physics, including 
condensed matter physcis \cite{Osborne2002N,Osterloh2002A,Vidal2003L}, 
quantum field theory \cite{Calabrese2004,Casini2008H,Casini2009,Hertzberg2011,Hertzberg2012,Seki2014,Grignani2016S},
quantum information theory \cite{Bennett2000,Nielsen2000}, 
quantum gravity and black hole physics \cite{tHooft1985,Susskind1994U,Kabat1995,Solodukin2006,Fursaev2006,Raamsdonk2006,Raamsdonk2010Gr,
Czech2012K,Swingle2014,Balasubramanian2015C} 
and the AdS/CFT correspondence in string theory 
\cite{Ryu2006T,Ryu2006TJHEP,Faulkner2013,Faulkner2013L,Faulkner2014GH}.
Since the seminal works of Bekenstein \cite{Bekenstein1973,Bekenstein1974} and Hawking \cite{Hawking1975} on the theory of black holes, thermodynamics of black hole has been a major driving force behind  
extensive study towards understanding fundamental relationship between gravitation, quantum theory, 
thermodynamics, and information theory for more than four decades. The cornerstone of this exploration 
is the black hole entropy, also known as the Bekenstein-Hawking entropy which may count the quantum 
microstates of the black hole.

A great deal of effort has been made towards finding fundamental degrees of freedom
responsible for the black hole entropy \cite{Wald2001}, which may lead us  eventually to a consistent quantum theory of gravity. The entanglement entropy \cite{Bombelli1986,Srednicki1993} was naturally brought 
into prominence to explain the origin of black hole entropy by defining quantum field theories on 
bipartite spaces \cite{Solodukhin2011}: This is because the entanglement entropy is proportional to the area of boundary surface, by which the space is divided into subspaces, just as the Bekenstein-Hawking entropy is proportional to the area of black hole horizon. However, the entanglement entropies,
which evaluated by employing local quantum field theories, diverge intrinsically due to divergent 
behavior of the two-point correlation functions in the UV region as usual, in contrast to the Bekenstein-Hawking 
entropy which is finite. It has been suggested \cite{Callan1994W,Susskind1994U} 
that the entanglement entropy is the first quantum correction while the Bekenstein-Hawking entropy is the black hole entropy at classical level. Then, the UV divergence of the entanglement entropy may be coped with the renormalization of procedure of quantum field theory: The divergent quantum corrections to the entanglement entropy may be absorbed into the renormalization of Newton constant \cite{Susskind1994U,Fursaev1995S}. 
However, we may find a mismatch between the renormalization of the entanglement entropy and that of Newton constant \cite{Solodukhin1995,Barvinsky1996} in the presence of non-minimal coupling, if we compare the one-loop quantum corrections to the gravitational action with the divergent entanglement entropy: This is called the puzzle of non-minimal coupling. 

A more promising approach to the entropy of black hole may be offered by string theory \cite{Susskind1994U}. Because string theory is supposed to be UV finite, the corresponding entropy in string theory may be finite: String theory at tree level may produce a finite entanglement entropy for black hole and Newton constant at the same time. It is also worthwhile to note that the Bekenstein-Hawking entropies have been obtained \cite{Strominger1996V,Callan1996M,Horowitz1996S} by counting the microstates of the BPS solitons \cite{Cvetic1995Y,Youm1999}, describing certain classes of extremal and nearly extremal black holes. Thus, it is likely that string theory may provide essential clues to understand the black hole entropy
at a fundamental level.  

Recent progress on black hole entropy stems largely from the AdS/CFT correspondence \cite{Maldacena1998}. According to the AdS/CFT correspondence, a conformal field theory (CFT) defined on the boundary of AdS space would be equivalent to a theory of gravitation in the bulk of AdS space. 
Along the line of AdS/CFT correspondence, 
Ryu and Takayanagi \cite{Ryu2006T,Ryu2006TJHEP} proposed that if we choose a closed subspace $\Sigma$ on the spatial boundary of the AdS space to define the entanglement entropy with respect to $\Sigma$, the entropy may be determined by $A(\Gamma)$, the area of minimal surface $\Gamma$, in the bulk AdS space which shares the same boundary with $\Sigma$ and Newton constant $G_N$ in the gravitational theory, dual to the CFT: $S_{\Sigma} = A(\Gamma)/4G_N$, $\p \Sigma = \p \Gamma$. Note that this proposed relationship resembles the Bekenstein-Hawking entropy formula for a black hole. It has been asserted indeed that the entanglement entropy of the boundary may be explained as the Bekenstein-Hawking entropy by using the proposed relationship \cite{Azeyanagi2008}.     

Although there have been numerous studies on the Bekenstein-Hawking entropy and the entanglement 
entropy, it remains an open question what the fundamental degrees of freedom are, responsible for the 
Bekenstein-Hawking entropy of black hole. This problem may be framed better in string field theory,
which takes full degrees of freedom of string into account. 
In this Letter, we shall study the entanglement entropy for open bosonic strings on $Dp$-branes by 
using the covariant string field theory \cite{Lee1988,Lee2017d,Lee2017JKPS,Lee2017De}. 

$Dp$-branes are extended objects described by spatial $p$-dimensional hyperplanes, upon which open strings can end. We may consider multiple $Dp$-branes which are located on a single hyperplane and divide the 
hyperplane into two half-hyperplanes along one of the transverse directions on it. Then, we define the
string density matrix by using string field in the Fock space representation.      
The entanglement entropy for open strings can be evaluated by applying the replica trick. The resultant 
entanglement entropy is proportional to the area of $(p-1)$ dimensional boundary of the bipartite hyperplanes and divergent as expected.  
However, the leading divergences in the UV region as well as in the IR region are mainly due to 
contributions of open string tachyons, which may be absent in the supersymmetric string theory. Except for these 
divergences due to the tachyon contributions, the entanglement entropy for open bosonic string on $Dp$-branes is finite for $2 \le p \le d_{\text{critical}} -2=24$ and logarithmically divergent for $p =1, 25$.  
The infinite tower of massive excitations of a string may drastically change the UV as well as IR behaviors 
of the theory.

\section{Open String on $D$-branes and Density Matrix}

$Dp$-branes are described by spatial $p$-dimensional hyperplanes, upon which open strings can end:
The end points of open string coordinates, $X^\m$, $\m = 0, 1, \dots, p$ satisfy the Neumann condition
and the end points of open string coordinates, $X^i$, $i=p+1, \dots, d$ satisfy the Dirichlet condition
\begin{subequations}
\beq
\frac{\p X^\m}{\p \s} \Bigl\vert_{\s = 0, \, \pi} &=& 0, ~~~\text{for}~~~ \m =0, 1, \dots, p, \\
X^i \Bigl\vert_{\s = 0, \, \pi}  &=& 0, ~~~ \text{for} ~~~ i = p+1, \dots, d . 
\eeq 
\end{subequations}
The string coordinates $X^\m$, $\m = 0, 1, \dots, p$ are tangential to the $Dp$-brane
world-volume while the string coordinates $X^i$, $i=p+1, \dots, d=d_{\text{critical}}-1$, are normal.
We consider $Dp$-branes with $1 \le p \le d_{\text{critical}} -1$.
In accordance with the boundary conditions,  
the string coordinates $X^I$, $I = 0, 1, \dots , d$ may be expanded in terms of normal modes as
\begin{subequations}
\beq
X^\m (\s) &=& x^\m + \sqrt{2} \sum_{n=1} x^\m_n \cos \left(n \s\right), ~~~ \m = 0, 1, \dots, p, \label{xmodem} \\
X^i (\s) &=& \sqrt{2} \sum_{n=1} x^i_n \sin \left(n \s\right), ~~~ i = p+1, \dots, d. \label{xmodei}
\eeq
\end{subequations}
Note that the string coordinates $X^i$, $i=p+1, \dots, d$ do not contain zero modes.
If the end points of the open strings are attached on $N$ multiple $Dp$-branes, the open string field
$\Psi$ carries the group indices of $U(N)$
\beq
\Psi [X] &=& \frac{1}{\sqrt{2}} \Psi^0[X] + \Psi^a[X] T^a, ~~~ a =1, \dots, N^2 -1
\eeq
where $T^a$ $a=1, \dots, N^2 -1$ are generators of $SU(N)$ group.

The BRST invariant string field theory action may be given as 
\beq
S_{BRST} &=& \int \text{tr} \left( \Psi * Q \Psi + \frac{2g}{3} \Psi * \Psi * \Psi \right),
\eeq 
by extending the Witten's cubic open string field theory \cite{Witten1986}. It takes only replacing usual 
normal mode expnsions of the string coordinates $X^I$ by those given as Eq.(\ref{xmodem}) and 
Eq.(\ref{xmodei}). Since we shall evaluate the entanglement entropy for free open string theory in this Letter, we take $g=0$. Integrating out the 
fermionic ghost zero modes, we find that the free string field action 
further reduces to the string field action in the proper-time gauge \cite{Lee1988} 
\begin{subequations}
\beq
S_0 &=& \int \tr\, \Psi \left(L_0 + L^{\text{gh}}_0 \right) \Psi, \label{freeaction}\\
L_0 + L^{\text{gh}}_0 &=& p^\m p^\n \eta_{\m\n}+  \sum_{n=1} n\, a^{\dag I}_n a^{J}_n \eta_{IJ} 
+\sum_{i=1}^2 \sum_{n=1} n a^{\text{gh}\dag}_{in} a^{\text{gh}}_{in} - 1 
\eeq 
\end{subequations}
where $a^{\text{gh}}_{in}$, $i=1,2$ are Fourier components of the BRST ghost coordinates. They satisfy,
$\{a^{\text{gh}}_{in}, a^{\dag \text{gh}}_{jm} \} = \d_{ij} \d_{nm}$. 
The usual BRST ghost coordinates may be expanded in terms of $a^{\text{gh}}_{in}$, $i=1,2$ as 
\begin{subequations}
\beq
b_{zz}(\s) &=& \frac{b_0}{2} + \frac{1}{2} \sum_{n=1} \left(a_{1n}^{\text{gh}}e^{-in\s} -i a^{\dag\text{gh}}_{2n} e^{in\s}\right), \\
b_{\bar z\bar z}(\s) &=& \frac{b_0}{2} + \frac{1}{2} \sum_{n=1} \left(a_{1n}^{\text{gh}} e^{in\s} -i a^{\dag\text{gh}}_{2n} e^{-in\s} \right), \\
c^z(\s) &=& \frac{c_0}{2} + \frac{1}{2}\sum_{n=1} \left(a^{\dag\text{gh}}_{1n} e^{in\s} + ia_{2n}^{\text{gh}} e^{-in\s}\right), \\
c^{\bar z}(\s) &=& \frac{c_0}{2} + \frac{1}{2}\sum_{n=1} \left(a^{\dag\text{gh}}_{1n} e^{-in\s} 
+ ia_{2n}^{\text{gh}} e^{in\s}\right).
\eeq
\end{subequations}

In order to evaluate the entanglement entropy we need to find well-defined local field operators representing open string. The Fock space representaion of the string field would be suitable for this
purpose:
\beq
\vert \Psi \rangle &=& \sum_{\{N^B_n, N^{\text{gh}}_n, n = 1, 2, 3, \dots\}} \sum_a \Psi^a_{\{N^B_n, N^{\text{gh}}_n\}}(x^\m) T^a \vert \{N^B_n, N^{\text{gh}}_n, n = 1, 2, 3, \dots \} \rangle \nn\\
&=& \sum_a \left(\phi^a(x) + A_\m^a(x) a^{\m\dag}_1 + \varphi^a_i(x) a^{i\dag}_1+ \dots \right) T^a\vert 0 \rangle. 
\eeq 
Here $\phi^a$, $A^a_\m, \varphi^a$ $a = 0, 1, \dots, N^2-1$ correspond to tachyon field, Yang-Mills gauge field, and massless vector fields respectively. 
The infinite tower of massive higher spin fields is abbreviated. 
Then, we find that the string vacuum wave functional may be written as follows 
\beq
\Phi[\Psi] &=& \langle 0 \vert \Psi\rangle = 
\frac{1}{Z} \int_{\Phi_{\{N_n, N^{\text{gh}}_n\}}( -\infty, x^1, \dots, x^p) = 0}^{\Phi_{\{N_n, N^{\text{gh}}_n\}}(0, x^1, \dots, x^p) = \Psi_{\{N_n, N^{\text{gh}}_n\}}(x^1, \dots,x^p)} D[\Phi] e^{-S_0(\Phi)}
\eeq
where ${Z}$ is a normalization constant. The vacuum density matrix for string field may be constructed in this basis as 
\beq
\rho[\Psi, \Psi^\prime] = \langle \Psi\vert 0 \rangle \langle 0 \vert \Psi^\prime \rangle = \Phi[\Psi]^* \Phi[\Psi^\prime] 
\eeq
in a similar way we define the vacuum density matrix for quantum field.

\begin{figure}[htbp]
   \begin {center}
    \epsfxsize=0.6\hsize

	\epsfbox{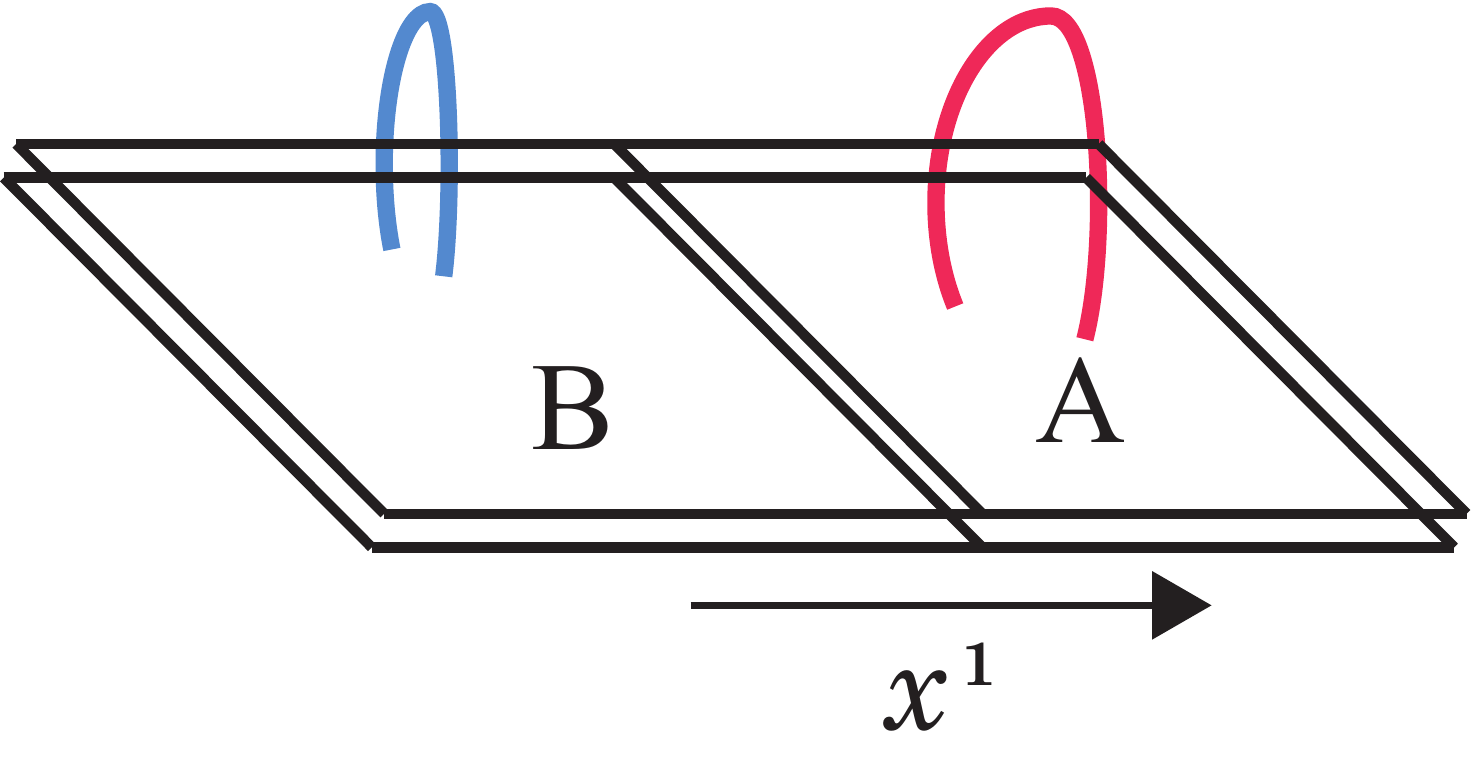}
   \end {center}
   \caption {\label{Openstring} Open strings on bipartite $Dp$-branes.}
\end{figure}

Now, we divide the $p$-dimensional hyperplane into two halves to make the 
half-hyperplane $A$ with $x^1 >0$ and the half-hyperplane $B$ with $x^1 <0$ as depicted by Fig.\ref{Openstring}.
It follows that string wave functional $\Psi$ may be written as a direct sum: $\Psi = \Psi_A \oplus \Psi_B$. 
Let us consider two wave functionals $\Psi = \Psi_A \oplus \Psi_B$ and $\Psi^\prime = \Psi^\prime_A \oplus \Psi_B$ which coincide on $B$. The reduced density matix on the half-hyperplane $A$ may be also defined in a 
similar fashion to the vacuum string wave functional. Summing over all possible functional $\Psi_B$, we have:
\beq
\rho_A \left(\Psi, \Psi^\prime\right) 
&=& \int D[\Phi_B] \Phi[\Psi_A \oplus \Psi_B]^* \Phi[\Psi^\prime_A \oplus \Psi_B] \nn\\
&=& \frac{1}{Z} \int_{\Phi_{\{N_n, N^{\text{gh}}_n\}}(0^-, x^1, \dots, x^p) = \Psi^\prime_{\{N_n, N^{\text{gh}}_n\}}(x^1, \dots,x^p)}^{\Phi_{\{N_n, N^{\text{gh}}_n\}}(0^+, x^1, \dots, x^p) = \Psi_{\{N_n, N^{\text{gh}}_n\}}(x^1, \dots,x^p)} D[\Phi] e^{-S_0(\Phi)}.
\eeq 
Consequetly, the $n$-th power of the reduced density matrix $\tr \rho^n_A$ may be expressed as 
a path integral over the $n$-sheeted Riemann surface
\beq
\tr \rho^n_A = \frac{Z_A(n)}{Z_A(1)^n}.
\eeq
Taking an analytic continuation of $n$ and differentiating with respect to $n$ at $n=1$, would 
yield the entanglement entropy for open string field 
\beq
S_{\text{Ent}} &=& - \frac{\p}{\p n} \, \tr \rho^n_A \Bigl\vert_{n=1} \nn\\
&=& - \frac{\partial}{\partial n} \Bigl\{\log Z(n) -n \log Z(1) \Bigr\}\Bigl\vert_{n=1}. 
\eeq 

\section{Evaluation of Entanglement Entropy for open string field}
The open string field in the Fock space representation is comprised of tachyons, massless fields, and infinite tower of massive fields. It is noteworthy that in the Fock space representation the free field actions Eq. (\ref{freeaction}) 
for component fields of open string are the same as those of scalar fields with masses given by eigenvalues of number operators $N_B + N_{gh}-1$:
\begin{subequations}
\beq
S_0 &=& \int  \sum_{\{N^B_n, N^{\rm{gh}}_n, n = 1, 2, 3, \dots\}} \Psi^{a\dag}_{\{N^B_n, N^{\rm{gh}}_n\}} \left(p^2 + N_B + N_{gh}-1 \right) 
\Psi^a_{\{N^B_n, N^{\rm{gh}}_n\}}, \\
N_B &=& \sum_{n=1} n a^{I\dag}_n a^I_n, ~~~ I = 0, 1, \dots, d  ,\\
N_{gh} &=& \sum_{n=1} n\left(a^{\dag\text{gh}}_{1n} a_{1n}^{\text{gh}} + a^{\dag\text{gh}}_{2n} a_{2n}^{\text{gh}} \right) . 
\eeq 
\end{subequations}
The statistics of the component field, $\Psi^a_{\{N^B_n, N^{\rm{gh}}_n\}}$ is determined by the total ghost number which it carries, $F_{gh}$
\beq
F_{gh} = \sum_{n=1}\left(a^{\dag\text{gh}}_{1n} a_{1n}^{\text{gh}} + a^{\dag\text{gh}}_{2n} a_{2n}^{\text{gh}} \right). 
\eeq
Thus, the entanglement entropy for open string can be obtained by directly extending the result of massive scalar field.

We may find an evaluation of entanglement entropy for a massive field 
in the literature \cite{Calabrese2004}. We shall extend it to evaluate the open string field theory with some modifications. 
For a free scalar field with mass $m$ in $(p+1)$ dimensions, the partition function $Z(n)$ defined on the 
$n$-sheeted Riemann surface may be written as 
\beq
\ln Z(n) = - \frac{1}{2} \ln {\rm Det} \left[-\Delta + m^2\right].
\eeq 
As we analytically continue $n$, near $n=1$, the $n$-sheeted Riemann surface becomes a conical space 
with a deficit angle $\d=2\pi(1-n)$, ${\cal R}^n$. 
The partition function $Z(n)$ is related to the Green's for a massive field in the conical space,
$G_n(\bolx, \bolx^\prime)$ as 
follows  
\beq
\frac{\partial }{\partial m^2} \ln Z(n) = - \half \int_{{\cal R}^n} d^{p+1} \bolx\,
\lim_{\bolx^\prime \rightarrow \bolx} G_n(\bolx, \bolx^\prime)  
\eeq 
where $\bolx = (x^0, x^1, \dots, x^p)=(x^0, x^1, \bolx_{\perp})$. 

An explicit expression 
of the Green's function $G_n(\bolx, \bolx^\prime)$, may be written by
\beq
G_n(\bolx, \bolx^\prime) = \frac{1}{2\pi n} \int \frac{d^{p-1} \bolp_\perp}{(2\pi)^{p-1}}\sum_{k=0}^\infty d_k \, \int^\infty_0 dq\,
q \frac{J_{k/n}(qr) J_{k/n}(qr^\prime)}{q^2+m^2+p^2_{\perp}} \cos \left(\frac{k}{n}(\th-\th^\prime)\right) e^{i\bolp_\perp \cdot (\bolx_\perp - 
\bolx^\prime_\perp)}
\eeq 
where $J$ is the Bessel function of the first kind and $d_0 =1$, $d_k = 2$ for $k\ge 1$. Here 
$(r, \th)$ is the polar coordinates on the two dimensional $(x^0,x^1)$-plane \cite{Calabrese2004}
and $\bolp_\perp = (p^2, \dots, p^p)$. Making use of the expicit expression of the Green's function 
in the coincident limit, $\bolx^\prime \rightarrow \bolx$, we get
\beq
\frac{\p}{\p m^2} \ln \tr \,\rho^n &=&  \frac{\p}{\p m^2} \ln \frac{Z(n)}{Z(1)^n} \nn\\
&=& - \half \left\{\int_{C^n} d^{p+1} \bolx G_n(\bolx, \bolx) - n \int d^{p+1} \bolx G_1(0)\right\}\nn\\
&=& - \frac{1 -n^2}{24n} A_\perp 
\int \frac{d^{p-1} \bolp_\perp}{(2\pi)^{p-1}} \frac{1}{m^2 + p^2_\perp}
\eeq
where $A_\perp$ is the area of boundary hypersurface, perpendicular to $(x^0,x^1)$-plane, 
$A_\perp = \int d^{p-1}\bolx_\perp$.  
Performing derivation with respect to $n$ and integrating over $m^2$ leads us to the entanglement 
entropy of a massive field 
\beq
S_A 
&=& \frac{A_\perp}{12} \int \frac{ds}{s} \int \frac{d^{p-1} \bolp_\perp}{(2\pi)^{p-1}} \exp\left\{
-s\left(p^2_\perp + m^2\right)\right\} \nn\\
&=& \frac{A_\perp}{12} \frac{1}{\left(8\pi^2\right)^{\frac{p-1}{2}}} \int_0^\infty dt \frac{1}{t^{\frac{p+1}{2}}} e^{-2\pi m^2 t} \label{scalar}
\eeq  
where $t=s/(2\pi)$. 

We may apply this result to the string field in the Fock space representation. A direct extension may yield the entanglement entropy of open string in the boson sector ($N^{\text{gh}}_n = 0, ~ n = 1, 2, \dots$):
\beq
S^{\rm Open}_A &=& \frac{A_\perp}{12} \int \frac{ds}{s} \int \frac{d^{p-1} \bolp_\perp}{(2\pi)^{p-1}}\, {\rm Tr}
\exp\left\{-s\left(p^2_\perp + N_B -1 \right)\right\} 
\eeq 
where `Tr' denotes the trace over the Fock space as well as over $U(N)$ group space.
It follows from bosonic harmonic oscillator algebra \cite{GreenSW} that 
\beq
{\rm Tr}\, e^{-sN_B} &=& N^2 \sum_{\{N^B_n\}} \exp \left\{-s\sum_{n=1} n N^B_n \right\} 
= N^2 \prod_{n=1} \left(\frac{1}{1-e^{-sn}}\right)^{d+1} = N^2 e^{-\frac{d+1}{24} s} \frac{1}{\eta\left(\frac{is}{2\pi}\right)^{d+1}}
\eeq 
where $\eta(\t)$ is the Dedekind eta-function defined by 
\beq
\eta(\t) = e^{i\pi \t/12} \prod_{n=1} \left(1- e^{2\pi i n\t} \right).
\eeq 
Thus, the entanglement entropy for open string in boson sector is evaluated as 
\beq
S_A^{\text{Open}} &=& \frac{A_\perp}{12} \frac{N^2}{\left(8\pi^2\right)^{\frac{p-1}{2}}} \int_0^\infty \frac{dt}{t}
\frac{1}{t^{\frac{p-1}{2}}} \exp\left\{\left(1- \frac{d+1}{24}\right) 2\pi t
\right\} \frac{1}{\eta(it)^{d+1}}  
\eeq 
where $t=s/(2\pi)$. 
 
Taking into account contributions of ghost sector, we may write the entanglement entropy for open string by
\beq \label{openbrst}
S^{\rm Open}_A &=& \frac{A_\perp}{12} \int \frac{ds}{s} \int \frac{d^{p-1} \bolp_\perp}{(2\pi)^{p-1}}\, {\rm Tr}
\exp\left\{-s\left(p^2_\perp + N_B + N_{gh} -1 \right)\right\}(-1)^{{F}_{gh}}
\eeq 
where 
\beq
N_{gh} = \sum_{n=1} n\left(a^{\dag\text{gh}}_{1n} a_{1n}^{\text{gh}} + a^{\dag\text{gh}}_{2n} a_{2n}^{\text{gh}} \right), ~~~
F_{gh} = \sum_{n=1}\left(a^{\dag\text{gh}}_{1n} a_{1n}^{\text{gh}} + a^{\dag\text{gh}}_{2n} a_{2n}^{\text{gh}} \right). 
\eeq 
We introduce the factor $(-1)^{{F}_{gh}}$ in Eq. (\ref{openbrst}) to take care of the fermion statistics of the ghost coordinates. 
Using 
\beq
\sum_{\{N_{gh}\}} e^{-s N_{gh}} (-1)^{{F}_{gh}} = \prod_{n=1} \left(1 - e^{-sn} \right)^2 
= e^{\frac{s}{12}} \eta\left(\frac{is}{2\pi}\right)^2 ,
\eeq
we find that the entanglement entropy $S^{\text{Open}}_A$ may be written as 
\beq
S^{\rm Open}_A &=& \frac{A_\perp}{12} \frac{N^2}{\left(8\pi^2\right)^{\frac{p-1}{2}}} \int_0^\infty \frac{dt}{t}
\frac{1}{t^{\frac{p-1}{2}}} \exp\left\{\left(\frac{25-d}{24}\right) 2\pi t
\right\} \frac{1}{\eta(it)^{d-1}} . 
\eeq 
For the bosonic string theory with $d=d_{crtical}-1 = 25$, 
\beq \label{Sopen}
S^{\rm Open}_A &=& \frac{A_\perp}{12} \frac{N^2}{\left(8\pi^2\right)^{\frac{p-1}{2}}} \int_0^\infty \frac{dt}{t} \frac{1}{t^{\frac{p-1}{2}}} \frac{1}{\eta(it)^{24}} . 
\eeq

The UV and IR behaivors of the integrand in Eq. (\ref{Sopen}) critically depend upon $p$. The IR behavor of the integrand may be read off from the asymptotic expansion of the Dedekind eta-function \cite{GreenSW}
\beq
\eta(it) = e^{- \frac{\pi}{12}t} \left( 1 - e^{-2\pi t} - e^{-4\pi t} + \dots \right), ~~~ t \rightarrow \infty.
\eeq 
In the asypmtotic region, the integrand becomes 
\beq \label{asymptotic}
\frac{1}{t^{\frac{p+1}{2}}} \frac{1}{\eta(it)^{24}} 
= \frac{1}{t^{\frac{p+1}{2}}}\Bigl\{e^{2\pi t} + 24 + {\cal O}(e^{-2\pi t}) \Bigr\} .
\eeq 
Camparing Eqs. (\ref{Sopen}, \ref{asymptotic}) with the entanglement entropy of a massive scalar 
Eq. (\ref{scalar}), we find that
the leading divergence is due to the tachyon contribution. 
Apart from the leading divergence, the 
entanglement entropy for open string is at most logarithmically divergent. Its IR behavor is 
finite for $ p \ge 2$ and logarithmically divergent for $p =1$. This result is consistent with our expectation that the 
infinite tower of massive states may improve the IR behavor of the intanglement entropy.  

The UV region corresponds to the region near $t=0$. Making use of the modular transformation formula for the 
Dedekind eta function 
\beq
\eta(-1/\t) = (-i\t)^{1/2} \eta(\t), 
\eeq 
and rewriting the integral for $S^{\text{Open}}_A$ in terms of $s = 1/t$, we get
\beq
S^{\rm Open}_A &=& 
 \frac{A_\perp}{12} \frac{N^2}{\left(8\pi^2\right)^{\frac{p-1}{2}}}\int_0^\infty ds\, 
s^{\frac{1}{2}\left(p-27\right)} \frac{1}{\eta(is)^{24}}.
\eeq 
The integrand in UV region where $s \rightarrow \infty$, may be expanded asymptotically as 
\beq
s^{\frac{1}{2}\left(p-27\right)} \frac{1}{\eta(is)^{24}} &=& 
s^{\frac{1}{2}\left(p-27\right)} \Biggl\{e^{2\pi s} + 24 + {\cal O}(e^{-2\pi s})
\Biggr\}.
\eeq 
The leading divergence is again ascribed to the tachyon contribution in the ``closed string channel", which may be also absent 
in the supersymmetric string theories. Except for the tachyon contribution, the entanglement entropy 
is UV finite for $p \le 24$ and logarithmically divergent for $p=25$. 

\section{Conclusions}

We studied the entanglement entropy for open boson strings on multiple $Dp$-branes for $1 \le p \le 25$.
In order to define local field operators, we employ covariant string field theories and the 
Fock space representation of the open string wave functionals. The $p$-dimensional hyperplane which constitutes 
the spatial dimension of $Dp$-branes is divided into two halves.     
By directly extending the calculation of entanglement entropy of a massive field, found in the 
previous work \cite{Calabrese2004}, to that of open string field, we were able to evaluate the entanglement entropy of open 
strings. It only takes some simple modifications. The entropy is found to be proportional to the 
area of the boundary of the bipartite $p$-dimensional hyperplanes as expected. However, its UV and IR behaviors are entirely
different from those of local field theory. It is divergent in both UV and IR regions. 
However, these divergences are attributable to the tachyon contributions, which may be absent in supersymmetric theories. 
If the leading divergences due to tachyons are excluded, the entanglement entropy of open string behaves much better 
than its counterpart of local field theory. It is finite in both UV and IR 
regions for 
$2 \le p \le d_{\text{critical}} -2 = 24$ 
and at most logarithmically divergent for $p=1, 25$. The present work strongly indicates 
that the infinite tower of massive states of open string drastically 
affects the IR behavior as well as the UV behavior of the entanglement entropy.  

A few remarks are in order before we conclude this Letter. Evaluation of the entanglement entropy 
of supersymmetric open strings on multiple $Dp$-branes may be an urgent and important task to be fulfilled 
in the near future. In the supersymmetric string theories, the entanglement entropies are expected to be 
finite or at most logarithmically divergent. 
Although the entanglement entropy of open string on $Dp$-branes is similar to the one-loop open string amplitude \cite{TLee2003VY},
the entanglement entropy of open string evaluated here is not a one loop correction but a tree level quantity.
Thus, this work supports the proposal of Susskind and Uglum \cite{Susskind1994U} that the black hole entropy may be understood as an entanglement entropy of string theory.
  
The massless gauge fields are contained in the string field as one of components. But the gauge symmetry is 
completely fixed in the BRST formulation. It suggestes that if we correctly take the BRST ghost 
sector into account when we evaluate the entanglement entropy, the component field $\Psi^a_{\{N^B_n, N^{\rm{gh}}_n\}}$ may be treated as a scalar field with a mass, $N_B + N_{gh}-1$. 
We only discussed the entanglement entropy of free open string field theory in this work. However, 
it may not be difficult to study the entanglement entropy of interacting open string field theory by extending 
the present work. The cubic interaction of open string field theory may produce classical and quantum corrections to the entanglement entropy.   

We may also study the entanglement entropies of open strings on more complex $Dp$-brane configurations,
which reduce to phenomenologically interesting field theoretical models in low energy limit.
Last but not least task may be to study the entanglement entropy in the
framework of covariant closed string field theory. 
The covariant closed string field theory in the proper-time gauge \cite{Lee2018EPJ} would serve as an essential tool to study the entanglement entropy of open string in this respect. 
In recent work, we have shown that the covariant string field theory in the proper-time gauge 
succesfully produces the scattering amplitudes of gravitons in the low energy limit. Comparison between  
the entanglement entropy of open string field and that of closed string field theory may shed some 
light on the related subjects. 

Note added. --- While writing the present Letter, I found a recent related work of Balasubramanian and Parrikar \cite{Balasubramanian2018}.
They studied the entanglement entropy of open string by using the light-cone field theory. Their result corresponds to the entanglement entropy of open string on a space-filling $D25$-brane. I also found that entanglement entropies in string theory have been studied in different settings in Refs. \cite{S.He2015,Zayas2015,Zayas2016}.

\vskip 1cm

\begin{acknowledgments}

TL was supported by Basic Science Research Program through the National Research Foundation of Korea(NRF) funded by the Ministry of Education (2017R1D1A1A02017805). This work is based on a talk given at the 2nd 
IBS-KIAS joint workshop at High 1, Korea (2018).

\end{acknowledgments}


\end{document}